\documentclass[preprint]{aastex63}
\usepackage[T1]{fontenc}
\usepackage[latin9]{inputenc}
\usepackage{bm}
\usepackage{amstext}
\usepackage{graphicx}

\makeatletter
\usepackage{savesym}
\savesymbol{iint}
\savesymbol{iiint}
\savesymbol{iiiint}
\usepackage{amsmath}
\usepackage{array}
\usepackage{multirow}
\usepackage{lineno}
\usepackage[caption=false]{subfig}

\makeatother

\usepackage{babel}
\begin{document}
\global\long\def\grad{\bm{\nabla}}%
\global\long\def\curl{\bm{\nabla}\times}%

\title{A spectral solver for solar inertial waves}
\author[0000-0001-6433-6038]{Jishnu Bhattacharya}
\affiliation{Center for Space Science, New York University Abu Dhabi, Abu Dhabi, P.O. Box 129188, UAE}
\author[0000-0003-2896-1471]{Shravan M. Hanasoge}
\affiliation{Department of Astronomy and Astrophysics, Tata Institute of Fundamental Research, Mumbai - 400005, India}
\affiliation{Center for Space Science, New York University Abu Dhabi, Abu Dhabi, P.O. Box 129188, UAE}

\begin{abstract}
    Inertial waves, which are dominantly driven by the Coriolis force, likely play an important role in solar dynamics, and additionally, provide a window into the solar subsurface. The latter allows us to infer properties that are inaccessible to the traditional technique of acoustic-wave helioseismology. Thus, a full characterization of these normal modes holds promise in enabling the investigation of solar subsurface dynamics. In this work, we develop a spectral eigenvalue solver to model the spectrum of inertial waves in the Sun. We model the solar convection zone as an anelastic medium, and solve for the normal modes of the momentum and energy equations. We demonstrate that the solver can reproduce the observed mode frequencies and line-widths well, not only of sectoral Rossby modes, but also the recently observed high-frequency inertial modes. In addition, we believe that the spectral solver is a useful contribution to the numerical methods on modeling inertial modes on the Sun.
\end{abstract}
\section{Introduction}

Measurements of oscillations in the Sun and stars provide a powerful means of constraining their interior structure and dynamics. For decades, p-mode seismology, where pressure is the restoring force for the oscillations, has been the focus of most efforts. However, recent discoveries of inertial modes of oscillations appear to have paved the way to a new form of seismology. Coriolis force and buoyancy-related effects are the restoring mechanisms for inertial modes, which, as a consequence, are differently sensitive than acoustic oscillations to interior rotation and structure parameters such as turbulent viscosity, the buoyancy frequency etc. Thus, inertial modes have the potential to provide altogether new constraints on aspects of the solar interior. Because the restoring mechanism is tightly connected to rotation, the associated frequencies of these modes are comparable to the rotation rate. 

Rossby waves, a class of inertial modes, have been widely studied in terrestrial \citep{pedlosky1987, pedlosky2003} and astrophysical settings \citep{2000ApJ...540.1102L, 2009A&A...493..193L, 2015ApJ...805L..14Z, 2021SSRv..217...15Z}. They have been observed prominently in the Sun recently, although their characteristics are somewhat unusual. The first such characteristic is that, despite the high degree of stratification and the presumably strong convective motion in the outer envelope as well as deeper inside, including distinct layers of rotational shear in both latitude and radius, the frequencies of the observed Rossby waves match closely with the canonical Rossby-Haurwitz dispersion relation derived for 2D \citep[see][]{2021SSRv..217...15Z}, $\omega = 2\Omega/(m+1)$, where $\omega/2\pi$ is the wave frequency, $\Omega$ the angular velocity of rotation, and $m$ is the azimuthal order. Secondly, only one branch of the Rossby-mode dispersion is observed, i.e., for $\ell = m$ in the Rossby-Haurwitz theory, where $\ell$ is spherical-harmonic degree. There is a whole range of lower frequencies, where $\ell > m$, that are predicted by the Rossby-Haurwitz theory but not (yet) observed in the Sun. Finally, only waves with vorticity that is symmetric around the equator are observed \citep{2018NatAs...2..568L, 2021A&A...652L...6G, 2022NatAs.tmp..102H}. The reasons for these exceptional behaviors are not known, although they may contain important insight into solar dynamics.

Another open problem is why thermal Rossby waves, so prominently predicted in numerical simulations, are not observed in the Sun. This may, of course, derive from aspects of solar structure that are not properly accounted for in numerical models.  \citet{1981A&A....94..126P, 1982ApJ...256..717S} formulated the equations governing Rossby waves in the Sun, but their formulations did not consider the extensive range of perturbations, such as entropy gradients and significant radial and latitudinal rotational shear. A more thorough study of the impact of these effects on inertial modes requires the analysis of a more general spectrum of the linearized Navier-Stokes equations. The low frequencies of oscillation ($\sim$0.5 $\mu$Hz or lower) suggests that inertial modes are decoupled from acoustic oscillations at much higher frequencies ($\sim$3000 $\mu$Hz). This separation of timescales allows us to invoke the anelastic approximation \citep{1969JAtS...26..448G, 1995GApFD..79....1B}, in which acoustic waves are filtered out and the time-derivative term of density in the mass-conservation equation is neglected. The complexity of the resultant equation, especially in the context of various additional perturbations, makes it resistant to purely analytical approaches.
The numerical study of the linearized Navier-Stokes equations in the anelastic limit for the Sun was first described in a seminal series of papers by \citet{1981ApJS...45..335G}, who focused on the onset of convection and the associated flow systems. \citet{1981ApJS...45..335G}, however, did not investigate the properties of Rossby waves in any great detail. Awaiting more detailed observations, this line of investigation has taken a back seat in the intervening decades.

Recent high-quality measurements of Rossby and inertial oscillation frequencies \citep{2018NatAs...2..568L, 2019A&A...626A...3L, 2019ApJ...871L..32H, 2020ApJ...891..125M, 2020A&A...634A..44P, 2021A&A...652L...6G, 2021A&A...652A..96M, 2022NatAs.tmp..102H}, have made this an opportune time to revisit these efforts. How to determine the relevant parameters and infer the underlying physics forms the focus of this article. In particular, we describe the equations and methodology that we apply to determine the spectrum of the anelastic equation. The use of spectral numerical techniques ensures high accuracy with a limited number of grid points. Anelasticity additionally reduces the dimensionality of the relevant set of eigenvalues and eigenfunctions and solutions.

A few comments are in order on how our method is similar to and distinct from existing approaches. Several numerical investigations into the spectrum of Rossby waves have been carried out recently, notably by \citet{2022A&A...662A..16B} and \citet{2022arXiv220413007T}.  The former set of authors uses a finite-difference approach, whereas the latter uses a spectral approach. The former solve for inertial and acoustic modes simultaneously, while the latter restrict themselves to the momentum equation for incompressible Rossby waves. Our work is similar to both in the sense that we solve an equation similar to that of the former, but it is similar to the latter because we use the anelastic approximation to eliminate acoustic oscillations and use a similar numerical scheme. Spectral approaches often permit an accurate estimation of the spectrum at lower resolutions than a finite-difference approach \citep{1988SJNA...25.1279W}, so this work aims to bring the analysis of inertial waves in the Sun closer to a form that can be used more accurately in forward or inverse problems at a reasonable computational expense. It also sheds possible light on the high-frequency inertial modes reported by \citet{2022NatAs.tmp..102H}.

\section{Method}
\subsection{Equation of motion}
We describe the motion of fluid in the solar interior as small perturbations about a hydrostatic reference state, characterized by the steady-state density $\bar{\rho}$, temperature $\bar{p}$ and pressure $\bar{p}$. We note that the knowledge of pressure and density specifies the temperature $\bar{T}$ through the equation of state, which, for an ideal gas, is given by $\bar{p}=\bar{\rho}R\bar{T}$ where $R$ is the ideal gas constant. We denote deviations to the reference state parameters by primed variables, e.g., $\rho^\prime$ represents a small fluctuation about the reference density $\bar{\rho}$. We do not apply a prime to the small-amplitude fluid velocity $\mathbf{u}$, since the reference state only contains axisymmetric rotation. Rotation distorts the spherically symmetric reference model of the Sun, as, in the rotating frame, the pressure gradient balances gravity as well as the centrifugal force \citep{1981A&A....94..126P}. In our analysis, we neglect this distortion, which limits the accuracy of our results to the first order in the angular frequency for slowly rotating stars.

\begin{figure*}
    \centering
    \includegraphics{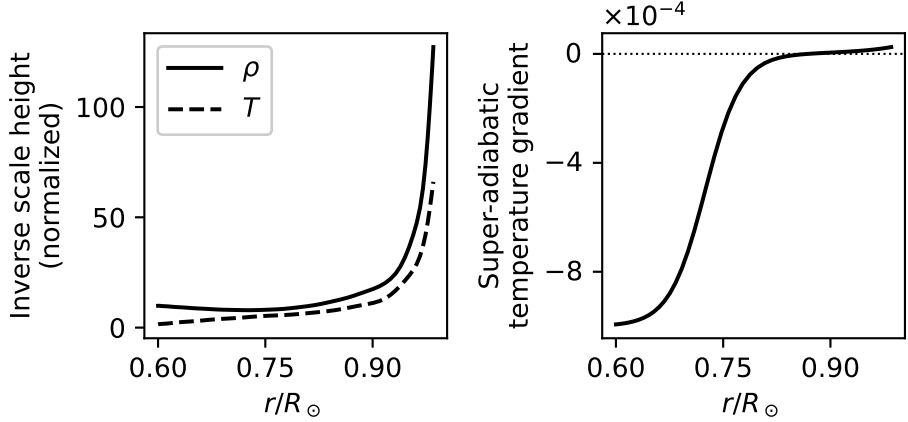}
    \caption{Left panel: Inverse scale heights for density $(\rho)$ and temperature $(T)$, multiplied by the solar radius. Right panel: Deviation from adiabatic temperature gradient $\delta(r)$.}
    \label{fig:scale_heights}
\end{figure*}

Small-amplitude oscillatory modes on the Sun occur over various widely separated time-scales. Acoustic and surface-gravity modes observed on the Sun have time-scales of minutes, whereas inertial modes oscillate over time-scales of months, corresponding to the solar rotation period. In our analysis, we filter out acoustic waves by applying the anelastic approximation to the equations of motion \citep{1981ApJS...45..335G, glatzmaier2014}. The key result that we use is that the equation of mass conservation may be expressed in the form
\begin{equation}
\grad\cdot\left(\bar{\rho}\mathbf{u}\right)=0.
\label{eq:anelastic_mass_conservation}
\end{equation} 
This implies that the divergence of velocity may be expressed as 
\begin{equation}
    \grad\cdot\mathbf{u}=\bar{\rho}\frac{d}{dr}\left(\frac{1}{\bar{\rho}}\right)u_{r}.
\end{equation}
We choose to represent this equation in terms of the negative inverse density scale height $\eta_\rho = d\left(\ln \bar{\rho}\right)/dr$ for notational convenience. The divergence of the velocity may therefore be expressed as $\grad\cdot\mathbf{u}=-\eta_{\rho}u_{r}$. We also define $D_{r\rho} = \partial_r + \eta_\rho$, which satisfies $\partial_r(\rho f(r))=\rho D_{r\rho}f(r)$ for an arbitrary radial function $f(r)$. Analogous to $\eta_\rho$, we may define the temperature scale $\eta_T=d\left(\ln \bar{T}\right)/dr$ that dictates the contribution to the stratification of the medium to energy diffusion. We plot the negative of the stratification functions $\eta_\rho$ and $\eta_T$ in Figure \ref{fig:scale_heights} (which are the inverse of the density and temperature scale heights) respectively. The sharp rise close to the surface indicates the high degree of stratification in near-surface layers.

Neglecting centrifugal forces, the time-evolution of the velocity $\mathbf{u}$ in a stratified, non-magnetic, convective solar interior is governed by the linearized Navier-Stokes equations, which may be expressed in the anelastic approximation as

\begin{align}
D_{t}\mathbf{u}=-2\boldsymbol{\Omega}\times\mathbf{u}-\grad\left(\frac{p^{\prime}}{\bar{\rho}}\right)-\frac{S^{\prime}}{c_{p}}\mathbf{g}+\frac{1}{\bar{\rho}}\mathbf{F}^{\nu},
\label{eq:eq_of_motion}
\end{align}
where $D_{t}=\partial_t + \mathbf{u}\cdot\grad$ is the material derivative in the rotating frame, $\mathbf{g}$ is the acceleration due to gravity, and $S^\prime$ is the entropy perturbation, and the viscous force $\mathbf{F}_\nu$ describes the loss of energy to smaller, unresolved length scales. In the assumption of constant kinematic viscosity and zero bulk viscosity, the viscous force $\mathbf{F}_\nu$ may be expressed as

\begin{align}
    \mathbf{F}_{\nu}=\nu\left[\grad\cdot\left(\bar{\rho}\left(\grad\mathbf{u}+\left(\grad\mathbf{u}\right)^{T}\right)\right)-\frac{2}{3}\grad\left(\bar{\rho}\grad\cdot\mathbf{u}\right)\right].
\end{align}

Equation \eqref{eq:eq_of_motion} includes the contribution of global-scale circulations such as meridional flows through the transport term $\mathbf{u}\cdot\grad$, but we have not included this in the present analysis. The impact of meridional flows on solar Rossby waves has been studied by \citet{2020A&A...642A.178G}. We have also assumed the Cowling approximation in dropping the perturbation to the acceleration due to fluctuations in the gravitational field.

We express the linearized energy equation in terms of the entropy $\bar{S}$, and its perturbation $S^\prime$, as 
\begin{align}
    \bar{\rho}\bar{T} D_t S^{\prime} & =\kappa\grad\cdot\left(\bar{\rho}\bar{T}\grad S^{\prime}\right),
    \label{eq:entropy_equation}
\end{align}
where $\kappa$ is the thermal conductivity. The anelastic approximation assumes an isentropic background medium --- in which case $\grad\bar{S}=0$ --- but convection in the Sun is driven by a mild entropy gradient in the convective envelope. The assumption built into our analysis is that, despite this gradient, the background medium remains close to isentropic. To model the solar convection zone, we include a radial entropy gradient arising due to a departure from adiabaticity:

\begin{equation}
    \frac{d\bar{S}}{dr}=\eta_\rho\gamma\delta,
\end{equation}
where $\delta$ is the super-adiabatic temperature gradient, and $\gamma$ is the adiabatic exponent that we choose as $1.64$. The value of $\delta$ is nearly zero in the convective envelope, since a small deviation from the adiabatic state is enough to drive large-scale convective instabilities, while in the radiative interior it takes the value $\delta\approx-0.1$. We model the profile of $\delta$ in the convention zone following \citet{2005ApJ...622.1320R}, such that it transitions from a positive value in the convection zone to asymptotically approach $-10^{-3}$ in the radiative zone. We plot the model of $\delta(r)$ that we have chosen in Figure \ref{fig:scale_heights}. The profiles of the thermal conductivity and viscosity are as yet unspecified. \citet{2021A&A...652L...6G} assume that the main source of dissipation is turbulence, implying that both the coefficients are of similar magnitudes. On such grounds, one might expect a radially varying conductivity, with the surface layers that harbor the strongest flows having a correspondingly high degree of diffusion, which drops rapidly towards the solar interior. Arguing along these lines, \citet{2014ApJ...789...35F} chose a profile that varies as $1/\sqrt{\rho}$. While such a model is physically more realistic, we disregard radial variations of conductivity and viscosity in the present work. However, we may include such a profile in our analysis, albeit at additional algebraic expense.

The set of equations \eqref{eq:anelastic_mass_conservation}, \eqref{eq:eq_of_motion} and \eqref{eq:entropy_equation} must be supplemented by appropriate boundary conditions on the velocity and entropy to complete the system. While the exact boundary condition at the solar surface is hard to pinpoint --- the surface being in vigorous motion with radiative transport --- we follow \citet{2011Icar..216..120J} and assume that the top and the bottom surfaces are impenetrable and stress-free, and that there is no entropy flux across the boundaries. These conditions may be expressed as

\begin{align}
    u_{r}=\partial_{r}\left(\frac{u_{\theta}}{r}\right)=\partial_{r}\left(\frac{u_{\phi}}{r}\right)=0,\quad\text{on}\;r=r_\text{in},\quad\text{and}\;r=r_\text{out},
    \label{eq:velocity_boundary_condition}
\end{align}
where $r_\text{in}$ and $r_\text{out}$ are the inner and outer radial extremities of the domain, respectively. The boundary condition on entropy may be expressed as
\begin{equation}
    \frac{dS^\prime}{dr} = 0,\quad\text{on}\;r=r_\text{in},\quad\text{and}\;r=r_\text{out}.
    \label{eq:entropy_boundary_condition}
\end{equation}

Aside from this, we also assume that all the quantities go to zero at the poles, such that the functions are single valued. In our analysis, we do not include the azimuthally invariant contribution to inertial waves, but it can be incorporated by adding appropriate boundary conditions at the poles.

\subsection{Uniformly rotating frame\label{sec:unirot}}

In a frame that is rotating at a constant angular velocity $\bm{\Omega}$, the material derivative $D_t\mathbf{u}$ of small velocity fluctuations is effectively equal to the partial time derivative $\partial_t\mathbf{u}$, as the quadratic term $\mathbf{u}\cdot\grad{\mathbf{u}}$ is much smaller in magnitude. We take the curl of Equation \eqref{eq:eq_of_motion} to eliminate the pressure perturbation term, and obtain the equation of motion for the vorticity to be

\begin{align}
\partial_{t}\curl\mathbf{u} & =2\left(\bm{\Omega}\cdot\grad\right)\mathbf{u}+\curl\left(\frac{1}{\bar{\rho}}\mathbf{F}_{\nu}\right)-\grad\left(\frac{S^{\prime}}{c_{p}}\right)\times\mathbf{g}.
\label{eq:eom_unirot}
\end{align}
The three vector components of Equation \eqref{eq:eom_unirot} are not all independent, as the vorticity $\curl\mathbf{u}$ is solenoidal. Without loss of generality, we may choose the independent equations that we work with to be the radial component of Equation \eqref{eq:eom_unirot}, and the radial component of a further curl of Equation \eqref{eq:eom_unirot}. We note that, in the anelastic approximation, we may obtain the pressure perturbation $p^\prime$ from the velocity $\mathbf{u}$ and the entropy perturbation $S^\prime$, and therefore do not lose information by eliminating it from our equations.

We proceed by switching to temporal frequency domain by assuming a harmonic time-dependence of $\exp(i\omega t)$ for the velocity, where the temporal frequency $\omega$ may be complex. The real part of the temporal frequency corresponds to oscillations, whereas the imaginary part signifies a decaying mode if positive, and a growing mode if negative. The choice of sign in the real part of the temporal frequency differs from conventional usage, where retrograde waves tend to have negative frequencies for positive $m$. In our analysis, the real part of the frequencies of retrograde modes are positive for $m>0$.

The anelastic velocity field has two independent components, and therefore it may equivalently be expressed in terms of two scalar stream functions. We may express the velocity as 

\begin{align}
    \mathbf{u}&=\bar{\rho}^{-1}[\grad\times\grad\times\left( \bar{\rho}\, W\left(r,\theta,\phi\right)\mathbf{e}_r\right)+\grad\times\left(\bar{\rho}\, V\left(r,\theta,\phi\right)\mathbf{e}_r\right)],
    \label{eq:velocity_stream_function}
\end{align}
where the radial and poloidal components of velocity may be derived from the stream function $W$, whereas the toroidal component may be derived from $V$.

We expand the stream functions on a basis of Chebyshev polynomials in radius and spherical harmonics in angular coordinates, as
\begin{align}
V\left(r,\theta,\phi\right) & =\sum_{\ell mq}V_{\ell mq}T_{q}\left(\bar{r}\right)\hat{P}_{\ell m}\left(\cos\theta\right)\exp\left(im\phi\right),\label{eq:V_expansion}\\
W\left(r,\theta,\phi\right) & =\sum_{\ell m q}W_{\ell mq}T_{q}\left(\bar{r}\right)\hat{P}_{\ell m}\left(\cos\theta\right)\exp\left(im\phi\right),\label{eq:W_expansion}
\end{align}
where $T_{q}$ represents the Chebyshev polynomial of degree $q$, $\bar{r}$ is the normalized radius defined as $\bar{r}=(r-r_\text{mid})/(\Delta r/2)$, where $\Delta r=r_\mathrm{out}-r_\mathrm{in}$ is the radial expanse of the domain, $r_\text{mid}$ is the midpoint of the radial span of the domain, and $\hat{P}_{\ell m}(\cos\theta)$ represents the normalized associated Legendre polynomial for an angular degree of $\ell$ and an azimuthal order $m$. Azimuthal symmetry of the background implies that Equation \eqref{eq:eom_unirot} decouples into individual equations for each $m$. In subsequent analysis, we suppress $m$ in the subscript wherever it is unambiguous, with the understanding that we solve the equations separately for each $m$. We also choose to refer to the field components for each $m$ with the subscript $m$, for example $V_m(r,\theta)=\sum_{\ell q}V_{\ell mq}T_{q}\left(\bar{r}\right)\hat{P}_{\ell m}\left(\cos\theta\right)$.

\subsection{Differential rotation}

It is well known that the Sun rotates differentially, with the fluid at the equator circulating faster than that at the poles. Waves in the Sun propagate through the spatially varying rotation field and are dragged along with the rotating fluid. We denote the angular velocity of rotation of the Sun by $\bm{\Omega}(r,\theta)$ and track the Sun in a frame that rotates at a uniform angular velocity $\bm{\Omega}_0$. We denote the differential rotation angular velocity by $\Delta\bm{\Omega}=\bm{\Omega}-\bm{\Omega}_0$. The angular velocities $\bm{\Omega}$ and $\bm{\Omega}_0$ are directed along the $\hat{z}$-axis, and we express the angular velocity of differential rotation as $\Delta\bm{\Omega}=\Delta\Omega\hat{z}$. The velocity of the fluid as measured in the tracking frame in this case may be expressed as a combination of the local rotation velocity $\mathbf{u}_{\Omega}=\Delta\bm{\Omega}\times\mathbf{r}=u_{\Omega\phi} \mathbf{e}_\phi$, and the intrinsic fluid velocity $\mathbf{u}_f$, as

\begin{align}
    \mathbf{u}=\mathbf{u}_f+\mathbf{u}_{\Omega}.
    \label{eq:u_intrinsic_rot}
\end{align}
The material derivative in this case contains a non-zero contribution at a linear order in $\mathbf{u}$ arising from the $\mathbf{u}\cdot\grad\mathbf{u}$ term. We rewrite this term as 

\begin{align}
    \mathbf{u}\cdot\grad\mathbf{u}=\frac{1}{2}\grad\left|\mathbf{u}\right|^{2}-\mathbf{u}\times\boldsymbol{\omega},
\end{align}
where $\boldsymbol{\omega}=\curl\mathbf{u}$ is the vorticity, which may be expressed using Equation \eqref{eq:u_intrinsic_rot} as
\begin{align}
    \boldsymbol{\omega}=\curl\mathbf{u}_f+\left[2\Delta\boldsymbol{\Omega}-\mathbf{r}\grad\cdot\Delta\boldsymbol{\Omega}+r\partial_{r}\left(\Delta\boldsymbol{\Omega}\right)\right].
\end{align}
We denote the intrinsic vorticity $\curl\mathbf{u}_f$ by $\bm{\omega}_f$, and the vorticity of differential rotation by $\bm{\omega}_\Omega$. The equation of motion, barring centrifugal terms, becomes 

\begin{align}
    \partial_{t}\mathbf{u}=\mathbf{u}\times\boldsymbol{\omega}-2\boldsymbol{\Omega}\times\mathbf{u}-\grad\left(\frac{p^{\prime}}{\bar{\rho}}+\frac{1}{2}\mathbf{u}^{2}\right)-\frac{S^{\prime}}{c_{p}}\mathbf{g}+\frac{1}{\bar{\rho}}\mathbf{F}^{\nu}.
    \label{eq:eom_diffrot}
\end{align}
The two extra terms $\mathbf{u}\times\grad\times\mathbf{u}$ and $\grad(\frac{1}{2}\mathbf{u}^{2})$, which were $\mathcal{O}(|\mathbf{u}|^2)$ in the uniform rotation case, now contribute to $\mathcal{O}(|\mathbf{u}|)$ through a coupling between the fluid velocity and differential rotation. The latter, however, does not contribute to the equation of motion for the vorticity, so the only extra term arising due to differential rotation is the former. We linearize the $\mathbf{u}\times\boldsymbol{\omega}$ term using Equation \eqref{eq:u_intrinsic_rot} to obtain
\begin{align}
\mathbf{u}\times\boldsymbol{\omega} & \approx\mathbf{u}_{f}\times\boldsymbol{\omega}_{\Omega}+\mathbf{u}_{\Omega}\times\boldsymbol{\omega}_{f}.
\label{eq:u_cross_omega}
\end{align}

Aside from its effect on the equation of motion, differential rotation also alters the entropy equation through a Doppler-shift term $-m\Delta\Omega S^\prime$, and a latitudinal transport term $(u_{\theta}/r)\partial_{\theta}\bar{S}$. An expression for $\partial_{\theta}\bar{S}$ may be obtained by invoking the Taylor-Proudman theorem \citep{2006ApJ...641..618M}. However, accounting for this term requires a resolution much higher than that we have access to. Since this term is proportional to $\cos\theta \partial_r \Omega - \left(\sin\theta/r\right) \partial_\theta \Omega$, and we restrict ourselves to equatorial modes where $(\partial_\theta \Omega)/\Omega_0$ is close to zero, we might expect the contribution of this term to not be substantial. Bearing this in mind, we don't include this term in our analysis. However, one would need to retain this term to study high-latitude modes.

Expressing the equations in this form makes it easier to carry out a spherical-harmonic decomposition. The situation simplifies significantly if the differential rotation is purely radial, without a latitudinal gradient. Such a profile is purely hypothetical, as the solar rotation rate varies considerably between the equator and the poles. Even so, it lets us pinpoint the contribution of the latitudinal variation of the rotation rate on the spectrum of inertial waves. In this work, we look at two rotation profiles: (a) a constant angular velocity $\Delta\Omega$, which, while not differential, implies that the Sun is rotating at a rate different from the tracking frame, (b) a radial profile $\Delta\Omega(r)$ of the angular velocity. Details of this radial variation are discussed in Section 3. It is possible to extend our analysis to latitudinal and radial solar-like rotation at additional algebraic expense.

\subsection{Equations for the stream functions}

We express Equations \eqref{eq:eq_of_motion} and \eqref{eq:entropy_equation} in terms of the stream functions $V_m$, $W_m$ and $S^\prime_m$ as an array of differential equations in $r$ --- one for each $\ell$. We normalize our fields to express them all in dimensions of velocity, choosing to work with the fields $V_m/R_\odot$, $i W_m/R_\odot^2$ and $\Omega R_\odot S^\prime_m/c_p$ instead, which enables us to compare their magnitudes. The equation for the associated-Legendre components of $V_m$ may be expressed in the form

\begin{align}
\frac{\omega}{\Omega}\left(\frac{V_{\ell}}{R_{\odot}}\right)&=\sum_{\ell^{\prime}}T_{VV,\ell\ell^{\prime}}\left(\frac{V_{\ell^{\prime}}}{R_{\odot}}\right)+T_{VW,\ell\ell^{\prime}}\left(\frac{iW_{\ell^{\prime}}}{R_{\odot}^{2}}\right),
\end{align}
where the operators $T_{VV}$ and $T_{VW}$ are given by 

\begin{align}
T_{VV,\ell\ell^{\prime}} & =\delta_{\ell\ell^{\prime}}\left[\frac{2m}{\ell\left(\ell+1\right)}-i E_\nu R_{\odot}^{2}\left(\partial_{r}^{2}-\frac{\ell\left(\ell+1\right)}{r^{2}}+\eta_{\rho}\left(\partial_{r}-\frac{2}{r}\right)\right)\right],\\
T_{VW,\ell\ell^{\prime}} & =-\frac{2}{\ell\left(\ell+1\right)}R_{\odot}\left[\ell^{\prime}\left(\ell^{\prime}+1\right)\left[\cos\theta\right]_{\ell\ell^{\prime}}\left(D_{r\rho}-\frac{2}{r}\right)+\left[\sin\theta\partial_{\theta}\right]_{\ell\ell^{\prime}}\left(D_{r\rho}-\frac{\ell^{\prime}\left(\ell^{\prime}+1\right)}{r}\right)\right],
\end{align}
where $E_\nu=\nu/\Omega R_{\odot}^{2}$ is the viscous Ekman number, and the matrix elements of the angular operators --- denoted by $A_{\ell \ell^\prime}$ --- are defined as 
\begin{equation}
A_{\ell\ell^{\prime},m}=\int_{0}^{\pi}d\theta\sin\theta \hat{P}_{\ell m}\left(\cos\theta\right)A\left(\theta\right)\hat{P}_{\ell^{\prime}m}\left(\cos\theta\right).
\end{equation}
We derive the explicit forms of the matrix elements of the angular operators in Appendix \ref{app:matrix_elements}.
Similarly, the equation for the associated-Legendre components of $W_m$ may be expressed as 
\begin{align}
\frac{\omega}{\Omega}B_{WW,\ell}\left(\frac{iW_{\ell}}{R_{\odot}^{2}}\right) & =\sum_{\ell^{\prime}}T_{WV,\ell\ell^{\prime}}\left(\frac{V_{\ell^{\prime}}}{R_{\odot}}\right)+T_{WW,\ell\ell^{\prime}}\left(\frac{iW_{\ell^{\prime}}}{R_{\odot}^{2}}\right)+T_{WW,\ell\ell^{\prime}}\left(\Omega R_{\odot}\frac{S_{\ell^{\prime}}^{\prime}}{c_{p}}\right),
\label{eq:W_equation}
\end{align}
where the operators are given by

\begin{align}
B_{WW,\ell} & =R_{\odot}^{2}\left[\partial_{r}D_{r\rho}-\frac{\ell\left(\ell+1\right)}{r^{2}}\right],\\
T_{WV,\ell\ell^{\prime}} & =-\frac{2R_{\odot}}{\ell\left(\ell+1\right)}\left(\ell^{\prime}\left(\ell^{\prime}+1\right)\left[\cos\theta\right]_{\ell\ell^{\prime}}\left(\partial_{r}-\frac{2}{r}\right)+\left[\sin\theta\partial_{\theta}\right]_{\ell\ell^{\prime}}\left(\partial_{r}-\frac{\ell^{\prime}\left(\ell^{\prime}+1\right)}{r}\right)\right),\\
T_{WW,\ell\ell^{\prime}} & =\delta_{\ell\ell^{\prime}}\left[R_{\odot}^{2}\,m\left(\frac{2}{\ell\left(\ell+1\right)}\left(\partial_{r}D_{r\rho}-\frac{\ell\left(\ell+1\right)}{r^{2}}\right)-\frac{2\eta_{\rho}}{r}\right)\right.\nonumber\\
 & -i E_\nu R_{\odot}^{4}\left\{ \left(\partial_{r}^{2}-\frac{\ell\left(\ell+1\right)}{r^{2}}\right)\left(\partial_{r}^{2}-\frac{\ell\left(\ell+1\right)}{r^{2}}+\frac{4}{r}\eta_{\rho}\right)\right.\nonumber\\
 & +\left(\partial_{r}-\frac{2}{r}\right)r\left(\partial_{r}^{2}-\frac{\ell\left(\ell+1\right)}{r^{2}}\right)\frac{\eta_{\rho}}{r}\nonumber\\
 & \left.\left.+\partial_{r}\left(\eta_{\rho}\left(\left(\partial_{r}-\frac{2}{r}\right)D_{r\rho}+\frac{\ell\left(\ell+1\right)}{r^{2}}\right)\right)-\frac{\ell\left(\ell+1\right)}{r^{2}}\left(2\eta_{\rho}\left(\partial_{r}-\frac{2}{r}\right)+\frac{2}{3}\eta_{\rho}^{2}\right)\right\} \right],\label{eq:Tww}\\
T_{WS,\ell\ell^{\prime}} & =-\delta_{\ell\ell^{\prime}}\frac{g\left(r\right)}{\Omega^{2}R_{\odot}}.
\end{align}

The terms in braces in $T_{WW,\ell\ell^\prime}$ represent the contribution of the viscous force, in which all terms aside from the first arise from the stratification of the medium. The operator $B_{WW,\ell}$ on the left-hand side is a key hurdle in casting the set of equations as a standard eigenvalue equation instead of a generalized one. An approach not explored in this work would involve inverting $B_{WW,\ell}$ to reduce the system to a standard eigenvalue problem, which would reduce the computational load significantly.

The entropy equation may be expressed as
\begin{align}
\frac{\omega}{\Omega}\left[\left(\Omega R_{\odot}\right)\frac{S_{\ell}^{\prime}}{c_{p}}\right] & =T_{SW,\ell\ell^{\prime}}\left(\frac{iW_{\ell^\prime}}{R_{\odot}^{2}}\right)+T_{SS,\ell\ell^{\prime}}\left[\left(\Omega R_{\odot}\right)\frac{S_{\ell^\prime}^{\prime}}{c_{p}}\right],
\label{eq:S_equation}
\end{align}

where the operators are given by

\begin{align}
T_{SW,\ell\ell^{\prime}} & =\delta_{\ell\ell^{\prime}}R_{\odot}^{3}\frac{\ell\left(\ell+1\right)}{r^{2}}\partial_{r}\left(\frac{\bar{S}}{c_{p}}\right),\\
T_{SS,\ell\ell^{\prime}} & =-\delta_{\ell\ell^{\prime}}\, iE_{\kappa}R_{\odot}^{2}\left[\partial_{r}^{2}+\frac{2}{r}\partial_{r}-\frac{\ell\left(\ell+1\right)}{r^{2}}+\left(\eta_{\rho}+\eta_{T}\right)\partial_{r}\right],
\end{align}
where $E_\kappa=\kappa/(\Omega R_{\odot}^{2})$ is the thermal Ekman number. Unlike the equations for the velocity stream functions, the diagonal operator in the entropy equation is purely diffusive, which indicates that modes where the entropy perturbation is the dominant term are primarily decaying or growing.

We translate the boundary conditions in Equations \eqref{eq:velocity_boundary_condition} and \eqref{eq:entropy_boundary_condition} to the stream functions to obtain independent radial constraints for each $\ell$:

\begin{align}
\sum_{q} V_{\ell q}\left(q^{2}- \frac{\Delta r}{r_{\text{out}}}\right) & =0,\label{eq:BC_V_1}\\
\sum_{q} \left(- 1\right)^{q} V_{\ell q}\left(q^{2}+ \frac{\Delta r}{r_{\text{in}}}\right) & =0,\label{eq:BC_V_2}\\
\sum_{q}\left(\pm 1\right)^{q} W_{\ell q} & =0,\label{eq:BC_W_12}\\
\sum_{q} \left(\pm 1\right)^{q} q^2 S^\prime_{\ell q} & =0,\label{eq:BC_S_1}
\end{align}
where the sum, in each case, is over the Chebyshev coefficients of the field. While these boundary conditions complete the system, the operator $T_{WW,\ell\ell^\prime}$ in Equation \eqref{eq:Tww} represents a fourth-order differential equation in radius, and we require additional boundary conditions to ensure uniqueness. We therefore supplement the system with additional zero-Neumann constraints on $W_\ell(r)$, requiring that its derivatives go to zero at the radial extremities. Experimentation shows that these additional constraints lead to smooth eigenvectors, while leaving the eigenvalues relatively unchanged (within $1\%$ for sectoral modes). We may express this Neumann constraint as

\begin{align}
    \sum_{q} \left(\pm 1\right)^{q} q^2 W_{\ell q} & =0.\label{eq:BC_W_34}
\end{align}
These boundary conditions are to be satisfied for each $\ell$, so a total of $n_\ell$ harmonic degrees leads to $8n_\ell$ constraints.

Differential rotation introduces additional terms in the equations. In this work, we look at a simplified scenario where the differential rotation rate depends solely on the radius. We define the fractional differential rotation rate $\Delta\tilde{\Omega}=\Delta\Omega/\Omega$, and evaluate the extra terms to be

\begin{align}
T_{D,VV,\ell\ell^{\prime}} & =\delta_{\ell\ell^{\prime}}m\Delta\tilde{\Omega}\left(\frac{2}{\ell\left(\ell+1\right)}-1\right),\\
T_{D,VW,\ell\ell^{\prime}} & =-\frac{2}{\ell\left(\ell+1\right)}R_{\odot}\left[\ell^{\prime}\left(\ell^{\prime}+1\right)\left[\cos\theta\right]_{\ell\ell^{\prime}}\left(\Delta\tilde{\Omega}\left(D_{r\rho}-\frac{2}{r}\right)-\frac{d}{dr}\left(\Delta\tilde{\Omega}\right)\right)\right.\nonumber\\
 & \left.+\left[\sin\theta\partial_{\theta}\right]_{\ell\ell^{\prime}}\left(\Delta\tilde{\Omega}\left(D_{r\rho}-\frac{\ell^{\prime}\left(\ell^{\prime}+1\right)}{r}\right)-\frac{\ell^{\prime}\left(\ell^{\prime}+1\right)}{2}\frac{d}{dr}\left(\Delta\tilde{\Omega}\right)\right)\right],\\
T_{D,WV,\ell\ell^{\prime}} & =-\frac{1}{\ell\left(\ell+1\right)}R_{\odot}\left[\left(4\ell^{\prime}\left(\ell^{\prime}+1\right)\left[\cos\theta\right]_{\ell\ell^{\prime}}+\left(\ell^{\prime}\left(\ell^{\prime}+1\right)+2\right)\left[\sin\theta\partial_{\theta}\right]_{\ell\ell^{\prime}}\right)\left(\frac{d}{dr}\left(\Delta\tilde{\Omega}\right)+\Delta\tilde{\Omega}\partial_{r}\right)\right.\nonumber\\
 & \left.+\left[\nabla_{h}^{2}\sin\theta\partial_{\theta}\right]_{\ell\ell^{\prime}}\left(\frac{d}{dr}\left(\Delta\tilde{\Omega}\right)+\Delta\tilde{\Omega}\left(\partial_{r}+\frac{2}{r}\right)\right)\right],\\
T_{D,VW,\ell\ell^{\prime}} & =\delta_{\ell\ell^{\prime}}mR_{\odot}^{2}\left[\left(\frac{2}{\ell\left(\ell+1\right)}-1\right)\left(\frac{d}{dr}\left(\Delta\tilde{\Omega}\right)D_{r\rho}+\Delta\tilde{\Omega}\left(\partial_{r}D_{r\rho}-\frac{\ell\left(\ell+1\right)}{r^{2}}\right)\right)\right.\nonumber\\
 & \left.-\Delta\tilde{\Omega}\frac{2\eta_{\rho}}{r}+\frac{d^{2}}{dr^{2}}\left(\Delta\tilde{\Omega}\right)+\frac{d}{dr}\left(\Delta\tilde{\Omega}\right)\left(\partial_{r}+\frac{2}{r}\right)\right],\\
T_{D,SS,\ell\ell^{\prime}} & =-\delta_{\ell\ell^{\prime}}m\Delta\tilde{\Omega},
\end{align}
where $\nabla^2_h$ is the lateral component of the Laplacian operator.

As a test case, we may consider the case of a constant $\Delta\Omega$, which corresponds to a uniformly rotating medium tracked at a different rate, in which case the terms simplify considerably.

\subsection{Numerical implementation}

We evaluate the spectrum of Rossby waves through an exact diagonalization of the matrix representation of the differential operator. The case where the rotation profile does not vary in the lateral directions is simpler to address, as the radial and angular terms are separated in variable, and each term appearing in the matrix may be represented as a Kronecker product of radial and angular operators. In general, this is not the case, e.g., the differential rotation velocity may not be separable in the radial and angular coordinates, and the matrices may need to be computed using a spectral or a pseudo-spectral approach in the angular coordinates.

In our analysis, we choose the coefficient of kinematic viscosity $\nu$ to be $2\times10^{12}\,\mathrm{cm}^2/\mathrm{s}$ (or equivalently, an Ekman number of $\nu/(\Omega_0 R_\odot^2)\approx1.45\times10^{-4}$), and set the thermal conductivity to be identical to the viscosity. We find that this value produces reasonable matches to the line-widths obtained by \cite{2020A&A...634A..44P}. This is not a best-fit estimate, but consistent with the choice made by \citet{2022A&A...662A..16B}, although a slightly lower value of $10^{12}\,\mathrm{cm}^2/\mathrm{s}$ had been suggested by \citet{2021A&A...652L...6G}. The actual situation regarding the values of the transport coefficients is indeed very different \citep{schu}, but what we have done is not an uncommon practice in numerical simulations of the Sun. We choose the radial domain of our analysis to correspond roughly to the solar convection zone; however, we set the lower boundary of the domain to $r=0.6R_\odot$ to adequately capture the sharp change in the adiabaticity profile at the base of the convection zone. We have verified that setting the lower boundary at $0.71R_\odot$ does not change the results appreciably. We place the outer boundary at $0.985R_\odot$ to avoid the steep stratification in the outermost layers of the Sun. 

In discretizing the operators, we follow \citet{2022arXiv220413007T} and split our analysis into two sets of modes: one for which the toroidal stream function $V$ is symmetric about the equator, and the other for which it is antisymmetric. In the symmetric case, only the spherical harmonic degrees $\ell=m,\;m+2,\;m+4\cdots$ contribute to $V$; whereas in the antisymmetric case, the contribution comes from $\ell=m+1,\;m+3,\;m+5\cdots$. We note that the stream function $W$ and the entropy perturbation $S$ have the opposite parity to $V$, so in the former case the spherical harmonic orders that contribute to $W$ are $\ell=m+1,\;m+3,\;m+5\cdots$, and the reverse in the latter. Such a separation effectively doubles the angular resolution of the eigenvalue problem. We also note that the toroidal stream function $V$ and the radial component of vorticity share the same angular profile, whereas the poloidal stream function $W$ shares its angular profile with the radial component of velocity. The arguments of equatorial symmetry, therefore, extend naturally to these fields as well.

We have developed a Julia implementation of the approach presented above. We choose the Julia language \citep{bezanson2017julia} as it is a high-performance, high-level language ideally suited to numerical applications. We make use of the freely available library ApproxFun.jl \citep{ApproxFun.jl-2014} to represent the radial operators as banded matrices (we note that similar functionality is provided in python by the Dedalus project \citep{2020PhRvR...2b3068B} and in Matlab by the library Chebfun \citep{Driscoll2014}, the latter being the inspiration for ApproxFun). The domain space of the operators are expanded in a basis of Chebyshev polynomials, and the range space in a basis of ultraspherical polynomials $U_q^\alpha(x)$ (which are also known as Gegenbauer polynomials), with the order $\alpha$ corresponding to the highest order of the radial derivative that appears in the equation. We use the term ``order'' in this context to denote the exponent $\alpha$ in the weight $(1-x^2)^{\alpha-1/2}$ that features in the definition of the inner product with respect to which the polynomials are orthogonal, and this differs from the degree $q$ of the polynomial. This approach is similar to that used by \citet{2022arXiv220413007T}. Such a sparse representation makes the evaluation of the operator matrices computationally inexpensive, and permits a purely spectral approach as opposed to a pseudo-spectral one. Owing to azimuthal symmetry, the angular operators may be represented as banded matrices in the basis of associated Legendre polynomials for a single $m$, and we have expanded on this in Appendix \ref{app:matrix_elements}. We use $60$ points in radial Chebyshev degrees, and $30$ in harmonic degree $\ell$, to obtain the discrete representations of the operators for each $m$. The matrix representations of self-adjoint radial operators thus obtained are not necessarily symmetric, and an approach similar to that used by \citet{2020JCoPh.41009383A} might improve the convergence of eigenvalues, although we have not explored this aspect in the present work.

We seek to solve the system
\begin{align}
    M \mathbf{x} &= \left(\frac{\omega}{\Omega}\right) B \mathbf{x} \label{eq:eigensystem}\\
    C \mathbf{x} &= 0, \label{eq:constraint}
\end{align}
where $M$ is the matrix representation of the restoring forces, $B$ contains the matrix representation of the double-curl, $\mathbf{x}=(V_{\ell m q},W_{\ell m q},S^\prime_{\ell m q})$ is the vector of the Chebyshev-Legendre basis components of the stream functions and the entropy perturbation, and $C$ is the matrix corresponding to the boundary conditions from Equations \eqref{eq:velocity_boundary_condition} and \eqref{eq:entropy_boundary_condition}. We start by computing the matrix $Z$ whose columns form a  basis for the null-space of $C$. We describe how we choose $Z$ in Appendix \ref{app:null_space}. By construction, an arbitrary vector ${\bf w}$ 
\begin{equation}
 \mathbf{x}=Z\mathbf{w}
 \label{eq:x_w_relation}
\end{equation}
satisfies the constraint in Equation \eqref{eq:constraint}. We rewrite the eigenvalue problem in Equations \eqref{eq:eigensystem} and \eqref{eq:constraint} as
\begin{align}
    \left(Z^T M Z\right) \mathbf{w} &= \left(\frac{\omega}{\Omega}\right) \left(Z^T B Z\right) \mathbf{w},
    \label{eq:transformed_eigensystem}
\end{align}
where $\mathbf{w}$ is unconstrained. We solve for the spectrum of eigenvalues and the corresponding eigenvectors $\mathbf{w}$ using LAPACK, and transform back to $\mathbf{x}$ using Equation \eqref{eq:x_w_relation}. Unlike \citet{2022arXiv220413007T} where the authors solve a sparse eigenvalue problem, we solve a dense one, which produces the full spectrum of eigenvalues and makes identifying spectral ridges easier. Unfortunately, our approach does not take advantage of the sparsity of the operator matrix.

We note that the matrix of operators that we thus construct is not Hermitian, which presents a significant challenge to the stability of solutions.  To obtain a set of eigenvectors stable to changing resolution, we perform a diagonal scaling of Equation \eqref{eq:transformed_eigensystem} as a preconditioning step, and rewrite the system as

\begin{align}
    \left[D_{1}\left(Z^{T}MZ\right)D_{2}^{-1}\right]\left(D_{2}\mathbf{w}\right)=\left(\frac{\omega}{\Omega}\right)\left[D_{1}\left(Z^{T}BZ\right)D_{2}^{-1}\right]\left(D_{2}\mathbf{w}\right).
\end{align}
Matrices $D_1$ and $D_2$ are block-diagonal, with the diagonal blocks equal to $\alpha_{i}\,I$, where $\alpha_{i}$ are real numbers chosen such that each block of $\left[D_{1}\left(Z^{T}MZ\right)D_{2}^{-1}\right]$ has an absolute maximum value of order $1$. While the choice of the matrices is carried out in a somewhat ad-hoc manner, we found that the solutions are insensitive to the exact choices once the maximum absolute values from all the blocks have the same order of magnitude. We note that such a scaling leaves the eigenvalues unchanged, and, following the diagonalization, we undo the scaling by left-multiplying the computed eigenvectors by $D_2^{-1}$.

Solutions to an eigenvalue equation in a Chebyshev basis are generally accurate if the eigenfunctions are smooth enough to be resolved on the Gauss-Chebyshev nodes \citep{1988SJNA...25.1279W}. Bearing this in mind, we impose the following filters to constrain the set of solutions:
\begin{itemize}
    \item Eigenfunctions must satisfy the boundary conditions to within a tolerance of $10^{-5}.$
    \item The original, unconstrained eigensystem (Equation \eqref{eq:eigensystem}) must be satisfied to within $0.01\%$.
    \item $90\%$ of the spectral power of the surface profile of the eigenfunction must lie within $\ell \le m+6$ where $\ell$ is the spherical harmonic degree.
    \item $90\%$ of the spectral power of the depth profile of the eigenfunction at the equator must lie within $n \le 6$, where $n$ is the degree of the Chebyshev polynomial.
    \item The imaginary parts of the eigenfrequencies must be non-negative, which eliminates growing modes.
    \item The angular profile of the eigenfunction at the surface must have its peak, as well as $30\%$ of the area under the curve within the latitudes of $\pm 30^\circ.$
\end{itemize}
These filters constrain us to decaying inertial mode solutions that have converged at the chosen resolution, and the exact cutoffs imposed are discretionary. We note that an equatorial filter precludes the study of high-latitude and critical-latitude modes that have been measured by \citep{2021A&A...652L...6G} and numerically studied by \citet{2022A&A...662A..16B}. Although such a restrictive filter is implemented here, this is not fundamental to our analysis, and this may be relaxed in future studies to investigate the categories of modes beyond those in this work.

The code has been made freely available on Github under the MIT license\footnote{https://github.com/jishnub/RossbyWaveSpectrum.jl}.

\section{Results}

\begin{figure}
    \centering
    \includegraphics[scale=0.7]{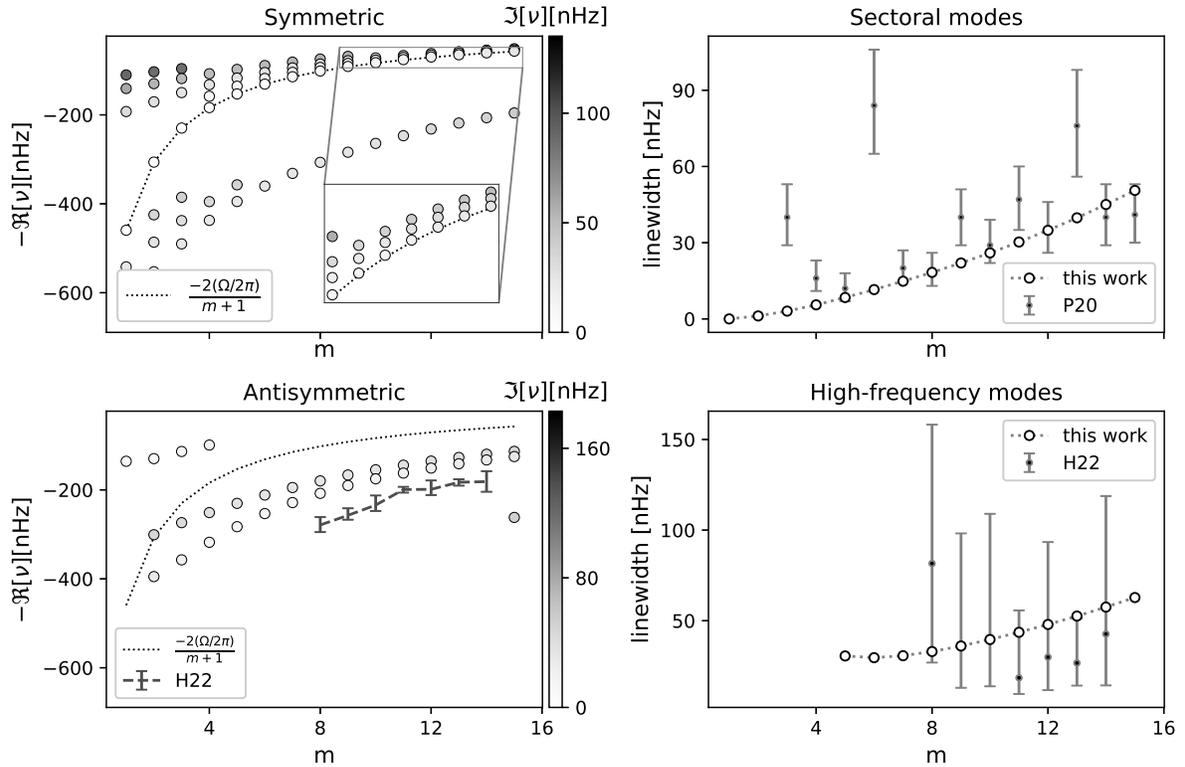}
    \caption{Top left: Spectrum of Rossby waves in a uniformly rotating Sun, for modes where the toroidal stream function $V(\mathbf{x})$ is symmetric about the equator. Bottom left: Same as top left, but for modes for which $V(\mathbf{x})$ is antisymmetric about the equator. The error bars indicate the high-frequency Rossby modes detected by \citet[][ring-diagram spectra]{2022NatAs.tmp..102H}. Top right: Line widths of sectoral, symmetric Rossby modes $(\omega\approx\frac{2\Omega}{m+1})$, as measured by \citet{2020A&A...634A..44P}, compared with our results. Bottom right: Line widths of high-frequency Rossby modes from observations \citep[][error bars]{2022NatAs.tmp..102H}, and our work (circles).}
    \label{fig:unirot_spectrum}
\end{figure}
\begin{figure*}
    \centering
    \includegraphics[scale=0.74]{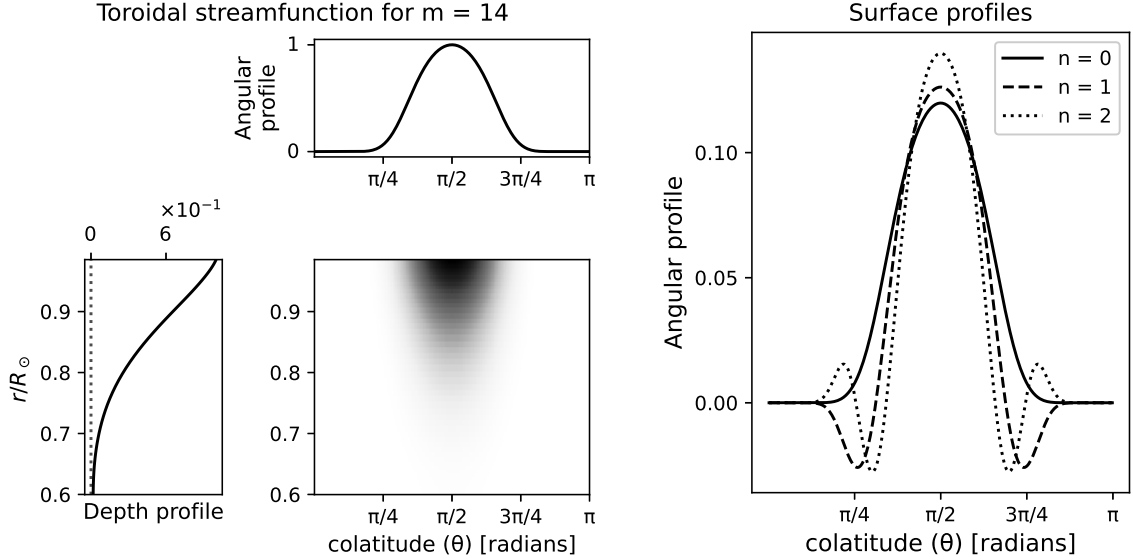}
    \caption{Left: Profile of the normalized toroidal stream function $V(\mathbf{x})$ for the sectoral mode for $m=14$, assuming that the Sun is rotating uniformly. The top panel is the angular profile at the surface, while the bottom left panel is the depth profile at the equator. Right: The angular profiles of the various toroidal stream functions $V_m(r,\theta)$ for $m=14$ at $r=r_\text{out}$. The index $n$ indicates the number of radial nodes.}
    \label{fig:sym_eigenfunction}
\end{figure*}

\begin{figure*}
    \centering
    \includegraphics[scale=0.5]{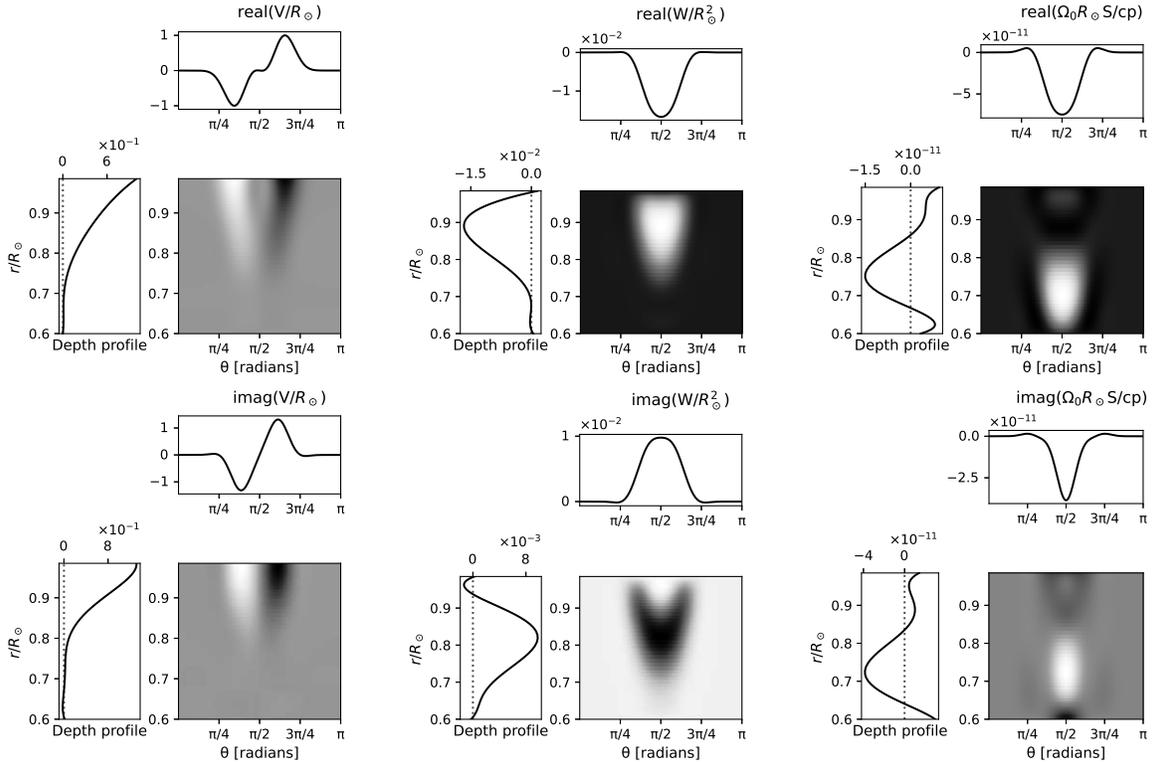}
    \caption{Antisymmetric high-frequency eigenfunction for $m=10$, having a peak frequency of $175$ nHz \citep[nearest the high-frequency ridge frequency as measured by][]{2022NatAs.tmp..102H}. The fields are all in units of velocity, albeit within an arbitrary overall normalization factor.}
    \label{fig:asym_eigenfunction}
\end{figure*}

\begin{figure*}
    \centering
    \includegraphics[scale=0.5]{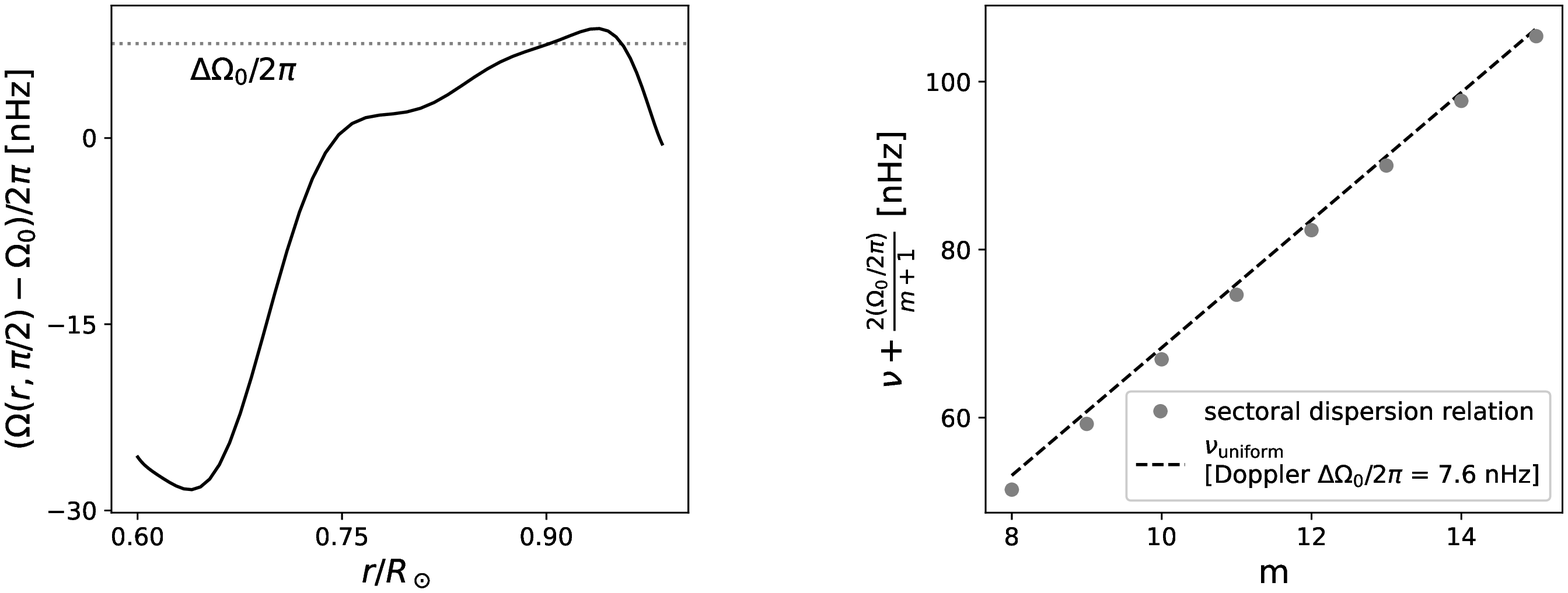}
    \caption{Left panel: A smoothed radial profile of differential rotation at the solar equator that we use in our analysis. Right panel: Doppler shifted dispersion relation of the sectoral modes (circles), and a least-square fit assuming a uniform rotation rate $\Omega_0+\Delta\Omega_0$, where $\Omega_0$ is the angular velocity of the tracking frame. The horizontal dotted line in the left panel indicates $\Delta\Omega_0/2\pi$, which is the best-fit uniform rotation rate obtained from this fit.}
    \label{fig:diffrot_rad_spectrum}
\end{figure*}

We plot the spectrum of Rossby waves in a uniformly rotating medium tracked in a frame rotating at the same rate in Figure \ref{fig:unirot_spectrum}, where the top and bottom left panels represent the spectrum obtained for symmetric and antisymmetric modes respectively. The dotted line represents the sectoral, thin-shell dispersion relation $\omega=2\Omega/(m+1)$. The modes that lie close to this relation contain most of their spectral power at $\ell=m$, and we refer to these as sectoral modes. In general, however, spectral power for a mode is spread across a range of harmonic degrees, as we might expect in a rotating medium that lacks spherical symmetry. The sectoral modes also have no radial nodes in the $V(\mathbf{x})$, or equivalently in the radial component of vorticity. We find several ridges below this analytical relation, corresponding to modes where the radial vorticity has increasing number of radial nodes. Interestingly, the spectrum also reveals a high-frequency ridge for modes with radial vorticity that is symmetric about the equator, with spectral power peaking at $\ell=m+2$. The top-right panel represents the line-widths of the latitudinally symmetric sectoral modes (corresponding to the dotted line in the top left panel). The bottom-left panel depicts latitudinally antisymmetric modes (denoted by circles), along with the frequencies of the high-frequency Rossby ridge as measured by \citet{2022NatAs.tmp..102H}. We find two distinct ridges that lie near the observed frequencies, but the modelled frequencies are lower than the observed ones. This is somewhat different from \citet{2022arXiv220413007T}, whose numerical frequencies are a little higher in absolute magnitude than the observed ones. There are various differences in the equations used in this work and that used by \citet{2022arXiv220413007T}, so such a disparity in the result is perhaps not surprising. We also recognize the possibility that, due to a lower resolution used in our work as compared to that used by \citet{2022arXiv220413007T}, some modes investigated by them do not appear in our spectrum. However, most of the power in the radial component of vorticity is concentrated at $\ell=m+1$ in the high-frequency modes that we obtain, which makes them resemble the observations by \citet{2022NatAs.tmp..102H}. It is therefore tempting to identify these as high-frequency Rossby-mode candidates. The bottom-left panel depicts the line widths of the high-frequency waves as obtained by us (circles), compared with that measured by \citet{2022NatAs.tmp..102H}. The line widths that we obtain are roughly consistent with the measured ones; as already remarked, this consistency might offer additional constraints on the coefficient of viscosity.

We plot the real part of the latitudinally symmetric toroidal stream function $V_m(r,\theta)$ for $m=14$ in Figure \ref{fig:sym_eigenfunction}. We also plot the stream functions for an antisymmetric, high-frequency mode for $m=10$ in Figure \ref{fig:asym_eigenfunction}. The toroidal stream function $V$ (and, as an extension, the radial vorticity) has no radial node, which makes them likely candidates for fundamental oscillation modes. The entropy perturbation peaks at the base of the convection zone, where there is a sharp change in the adiabatic index, whereas the velocity stream functions peak close to the solar surface. This result is somewhat in contrast to \citet{2022arXiv220413007T}, who find that there is a substantial poloidal component deep within the convection zone. It is unclear how strongly their result depends on their choice of incompressibility in the equation of continuity, which ignores the radial stratification of density. Further work might be necessary to reconcile their results with ours. Aside from the high-frequency ridge discussed above, a slightly lower-frequency ridge is visible in the bottom left subplot of Figure \ref{fig:unirot_spectrum}, which corresponds to modes having one radial node.

We plot the spectrum of inertial waves in a radially differentially rotating Sun in Figure \ref{fig:diffrot_rad_spectrum}. The rotation velocity in our model has the functional form corresponding to the subsurface profile at the solar equator. The left panel illustrates the rotation rate as a function of radius, while the right panel shows the dispersion relation for the sectoral Rossby modes (circles), and a best-fit dispersion relation assuming a constant shift in the angular velocity by $\Delta\Omega_0$ (dashed line). Not all eigenfunctions converge to smooth solutions at the resolution used in the latter case, and as a consequence, there are gaps in the spectral ridges. Unfortunately, increasing the resolution beyond this becomes challenging due to resource limitations, and alternate algorithms might be necessary to achieve convergence for these modes. The interesting observations in these are that the Doppler shift appears to push the modes from ones propagating in a retrograde sense to prograde ones. The switch occurs for an $m$ where the oscillation frequency $2(\Omega_0 + \Delta\Omega_0)/(m+1)$ becomes comparable to the Doppler shift $m\Delta\Omega_0$, assuming a constant shift $\Delta\Omega_0$. However, this frequency is specific to the tracking frame chosen and not intrinsic to the oscillation in the Sun, and additionally, this effected will be mitigated by latitudinal differential rotation, which --- owing to a reduction in the rotation rate with increasing latitude --- might shift the frequencies in the opposite sense. Further work is therefore necessary to establish the spectrum in the background of solar-like differential rotation that varies both in radius and latitude. Interestingly, the modes for $m>10$ closely follow the dispersion relation obtained for a constant shift in the rotation rate, whereas the lower $m$ modes depart from this relation as expected, as these modes extend deeper into the Sun. The best-fit angular velocity $\Delta\Omega_0$ therefore provides an estimate of the depth sensitivity of these waves.

\section{Conclusion}
We have described a spectral numerical technique by which to extract the eigenbasis of the linear anelastic operator, which corresponds to inertial waves in the Sun. We can reproduce the central frequencies and line widths of Rossby waves on the Sun, as measured by \citet{2020A&A...634A..44P}. Our results also qualitatively agree with that by \citet{2022arXiv220413007T} and support the identification of the high-frequency spectral ridge observed  by \citet{2022NatAs.tmp..102H} as inertial waves, although a uniformly rotating model of the Sun appears to produce somewhat lower oscillation frequencies. Further investigation into this in presence of solar-like differential rotation remains to be carried out. Interestingly, we also see a high-frequency ridge for equatorially symmetric radial vorticity, which, to our knowledge, has not been observed on the Sun.

The accuracy and generality of the method presented in this work offer promise for future investigations into solar structure and dynamics using measurements of inertial oscillations. In further work, we will focus on improving the computational expense of the approach, to render it feasible to use techniques such as Markov-chain Monte Carlo (MCMC) for determining the best-fit set of parameters that can explain the observed mode frequencies, in addition to allowing the estimation of proper uncertainties. The setup can be extended in a straightforward manner to include magnetism, since the Lorentz force term is very similar in structure to the form that differential rotation takes.  

\acknowledgments
This material is based upon work supported by Tamkeen under the NYU Abu Dhabi Research Institute grant G1502. We also acknowledge support from
the King Abdullah University of Science and Technology (KAUST) Office of Sponsored Research (OSR) under award OSR-CRG2020-4342.
This research was carried out on the High-Performance Computing resources at New York University Abu Dhabi.

\appendix
{}
\section{Matrix elements of angular operators\label{app:matrix_elements}}

We evaluate the matrix elements of $\cos\theta$ and $\sin\theta\partial_\theta$ in the basis of normalized associated Legendre polynomials. For an operator $A(\theta)$, the matrix elements in an associated Legendre basis are 
\begin{equation}
A_{\ell\ell^{\prime},m}=\int_{0}^{\pi}d\theta\sin\theta \hat{P}_{\ell m}\left(\cos\theta\right)A\left(\theta\right)\hat{P}_{\ell^{\prime}m}\left(\cos\theta\right).    
\end{equation}
In subsequent analysis, we suppress the subscript $m$, with the understanding that we compute the matrix for a specific azimuthal order.
We note that if the action of an operator $A(\theta)$ on $\hat{P}_{\ell m}(\cos\theta)$ is described by \begin{align}
    A(\theta)\hat{P}_{\ell m}\left(\cos\theta\right)=\sum_{\ell^\prime} C_{\ell\ell^{\prime}}\hat{P}_{\ell^{\prime}m}\left(\cos\theta\right),
\end{align}
the coefficients $C_{\ell\ell^{\prime}}$ represent the matrix elements of the transpose of $A$, that is, $A_{\ell\ell^{\prime}}=C_{\ell^{\prime}\ell}.$ We use the relations 

\begin{align}
    \cos\theta\hat{P}_{\ell m}\left(\cos\theta\right)&=\sqrt{\frac{\left(\ell-m+1\right)\left(\ell+m+1\right)}{\left(2\ell+1\right)\left(2\ell+3\right)}}\hat{P}_{\ell+1m}\left(\cos\theta\right)+\sqrt{\frac{\left(\ell-m\right)\left(\ell+m\right)}{\left(2\ell-1\right)\left(2\ell+1\right)}}\hat{P}_{\ell-1m}\left(\cos\theta\right),\\
    \sin\theta\partial_{\theta}\hat{P}_{\ell m}\left(\cos\theta\right)&=\ell\cos\theta\hat{P}_{\ell m}\left(\cos\theta\right)-\sqrt{\frac{2\ell+1}{2\ell-1}\left(\ell^{2}-m^{2}\right)}\hat{P}_{\ell-1m}\left(\cos\theta\right),
\end{align}
to evaluate the matrix elements of $\cos\theta$ and $\sin\theta\partial_\theta$ in the normalized associated Legendre polynomial basis. We may use these as building blocks to compute the matrix elements of operators that may be expressed as the product of these terms.

\section{Radial basis\label{app:null_space}}

Given the constraints in Equations \eqref{eq:BC_V_1}-\eqref{eq:BC_W_34}, we compute three sets of bases --- one for each of $V_\ell$, $W_\ell$ and $S^\prime_\ell$ --- that automatically satisfy the constraints, which enables us to expand the field in the corresponding basis. We use two different approaches to compute the basis. For $V_\ell$, given $n$ Chebyshev coefficients, we represent the constraints for each $\ell$ as a $2\times n$ matrix, and compute an orthogonal basis for its null space through a full singular-value decomposition \citep[see e.g.][]{Porcelli2015ASP}. The Julia standard library LinearAlgebra conveniently contains a function ``nullspace'' that provides an implementation of this algorithm.

The bases for $W$ and $S^\prime$ are easier to compute through a basis-recombination approach \citep{Heinrichs1989}. We define our basis in terms of the normalized radius $\bar{r}=(r-r_\mathrm{mid})/(\Delta r/2)$, which takes the values $\pm 1$ at the radial boundaries of the domain. The first basis that we choose is 
\begin{align}
    Q_{q}\left(\bar{r}\right) &= \left(1-\bar{r}^{2}\right)^2 T_{q}\left(\bar{r}\right),
\end{align}
which satisfies
\begin{equation}
    Q_{q}\left(\pm1\right)=Q_{q}^{\prime}\left(\pm1\right)=0.
\end{equation}
We may therefore expand $W_\ell(r)$ as 
\begin{equation}
    W_{\ell}\left(r\right)=\sum_{q}W_{\ell q}Q_{q}\left(\bar{r}\right).
\end{equation}
For the Neumann condition on $S^\prime(r)$, we choose the basis to be
\begin{equation}
    P_{q}\left(\bar{r}\right)=T_{q}\left(\bar{r}\right)-\frac{p^{2}}{\left(p+2\right)^{2}}T_{q+2}\left(\bar{r}\right).
\end{equation}
Using the result $T_q^\prime(\pm 1)= (\pm 1)^{q+1} q^2$, we obtain $P_{q}^{\prime}\left(\pm1\right)=0,$ so we may expand $S^\prime_\ell(r)$ as
\begin{equation}
    S_{\ell}^{\prime}\left(r\right)=\sum_{q}S_{\ell q}^{\prime}P_{q}\left(\bar{r}\right).
\end{equation}

Collectively, these three bases ensure that the functions $V_\ell(r)$, $W_\ell(r)$, and $S^\prime_\ell(r)$ satisfy the boundary conditions. We note that the choice of the bases is not unique, and the bases $Q_q$ and $P_q$ do not form orthogonal sets. However, orthogonality is not crucial, and a different choice of basis, for example the orthogonal basis obtained through the ``nullspace'' function, does not change the results, which enhances our confidence in the solutions. The choice made here leads to smooth basis elements, and, as a consequence, the results are easier to interpret in this basis.

\begin{figure*}
    \centering
    \includegraphics{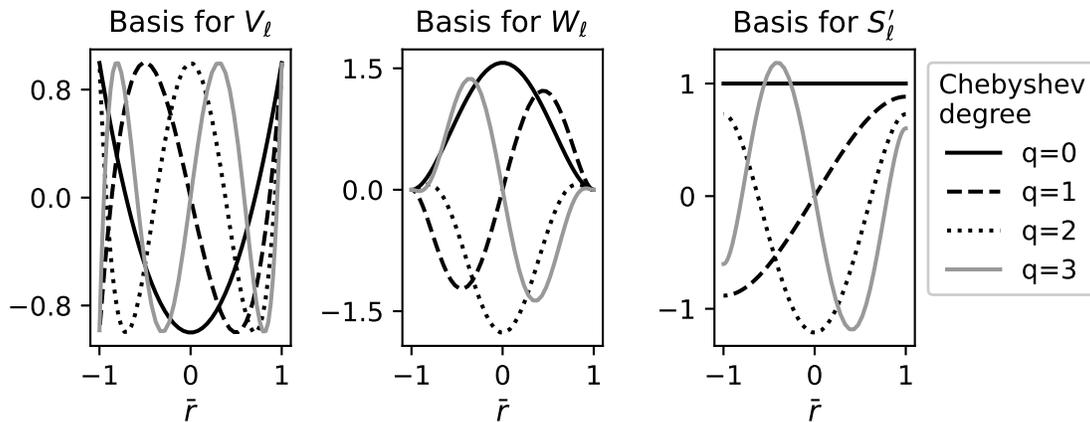}
    \caption{First few basis functions chosen to satisfy the boundary conditions on $V_\ell(r)$ (left panel), $W_\ell(r)$ (middle panel) and $S^\prime_\ell(r)$ (right panel).}
    \label{fig:basis}
\end{figure*}

We plot the first few basis functions for each of the three fields in Figure \ref{fig:basis}.

\bibliographystyle{aasjournal}
\bibliography{references}

\end{document}